\documentclass[aps,showpacs,twocolumn,superscriptaddress]{revtex4}
\usepackage{graphicx}
\usepackage{dcolumn}
\usepackage{amsmath}

\begin{document}

\title{ Non-deterministic random bit generator based on electronics noise }

\author{Mario Stip\v cevi\' c}
\email{Mario.Stipcevic@irb.hr}
\affiliation{\footnotesize Rudjer Bo\v{s}kovi\'{c} Institute,
         Bijeni\v cka 54, P.O.B. 180, HR-10002 Zagreb, Croatia}

\date{\today}

\begin{abstract}
\rule{0ex}{3ex}
\noindent
Non-deterministic random bits are needed in many scientific fields.
Unfortunately today's computers are very limited in ability to
produce them.
We present here 
a method for extraction of non-deterministic random bits from 
random physics processes, and one practical realization of a physical
generator based on it.
The method is shown to deliver increasingly good randomness in the limit 
of slow sampling.
A sample of approximately $10^9$ bits produced 
by the physical generator prototype
is subjected to a series of well-known
statistical tests showing no weaknesses.
\end{abstract}

\maketitle

\section{Motivation}
\

Today's computers are Turing machines governed by deterministic laws.
It has been noticed that such machines can not solve certain
class of problems because of their inability to produce random numbers.
For example, it is impossible to simulate a simple radioactive decay.
A computer program that would be able to do this must 
be able to produce
a sequence of times of decays which is by no means deterministic.
\\

Consequently, a complete machine that is believed to be 
the ultimate universal
computing machine is referred to as 
"Turing machine with a random number generator" or 
"probabilistic Turing machine" \cite{QCbook, Wikipedia-pTuring}.
\\

Since computers play an increasingly important role in modern science and life,
it is also increasingly important to develop "good" random number generators.
\\

The classical approach is to approximate truly random, non-deterministic
generator of random numbers by a carefully chosen mathematical
function which produces approximately random numbers.
Such a function of course can be calculated on Turing machines.
Without loss of generality we will only consider random bit generators (RBG).
The basic technique is as follows.
First, one chooses at will an integer number $x_0$ (so called {\em seed}) 
from a large
set of numbers that are known to be valid seeds for a given function $f$.
(Good coverage of the theory and practice of pseudo-random generators may be found in \cite{CRC, NRinC, Knuth, RANLUX, RANLUX2, Hellekalek, rreview, Miller, Marsag90}.)
Then one calculates:
\begin{eqnarray}
x_1 & = & f(x_0)
\end{eqnarray}

 Iterating this as follows:

\begin{eqnarray}
x_{i+i} & = & f(x_i), \,\,\, i = 1, 2, 3 \ldots
\end{eqnarray}

 results in a sequence of numbers: 

\begin{equation}
x_1, x_2, x_3, ...
\label{eq:sum}
\end{equation}

These numbers converted to bits and put together side-by-side 
form a long sequence of {\em pseudo-random bits}. Sometimes, not all
bits 
from the numbers $x_i$ are used to form the sequence but only a 
pre-selected part (for example the most significant half).
\\

Variants of this basic technique exists but the underlying property of 
all pseudo-random generators is that they {\em must} accept a seed, 
a form of mathematical 
initial state that completely determines the sequence of bits 
produced thereafter.
The idea behind all this is that to someone who does not know the seed
and/or function $f$ (or does not care about them), the sequence
of bits produced appears to be random. 
\\

Pseudo-random numbers can be produced efficiently and are 
used a lot in simulations of stochastic processes like passage of
particles through matter \cite{Geant} or speeding up of calculation of
exact problems such as multi-dimensional integration or
primality testing \cite{primality}.
\\

While good generator functions
which approximate true randomness quite well for most 
purposes \cite{RANLUX,RANLUX2,Hellekalek,Marsag90},
it is absolutely clear that the entropy of the
whole pseudo-random sequence, no matter how long, is equal
to the entropy of the seed only, which is usually not more
than 16 to 64 bits. This is most clearly seen by noting that the optimal
compression of a sequence of type (\ref{eq:sum}) would result 
only in the seed $x_0$ and
a decompression routine that equals $f$. Since $f$ is usually publicly 
known it does not represent a useful source of entropy, 
thus the only entropy left is that of the seed.
And even this tiny entropy, namely the entropy of the seed,
has to be
provided by somebody or something that has nothing to do
with the pseudo-random generator itself. Therefore the conclusion
that pseudo-random generators do not generate entropy
seems unavoidable. 
\\

For applications in cryptography this may be a killing property.
For example provability of information-theoretic security of 
Quantum cryptography protocols such as 
BB84 \cite{BB84}, 
or Maurer's SKAPD protocol \cite{Mau93}
assumes existence of generators of truly random
(non-deterministic) numbers. Even much less complicated applications
such as PIN or TAN number generators may not be proven secure if
the numbers generated are based on a deterministic procedure.
Cases are known where seeding by a low-entropy source (such as clock)
led to serious compromise of the subsequent cryptographic protocol.
An example is attack to the Netscape's 40 bit RC4-40
\cite{rc4}
challenge data and encryption keys which could
be revealed in a minute or so \cite{dobbs96}. 
The authors of that article suspect that 128 bit version RC4-128 
would not be much harder to break either if seeding is done in
a similar fashion.
(This example also shows how a perfectly good cryptographic method
may be ruined by a non-educated or malicious implementator.)
\\

All these cases boil down to the fact that
it is hard to get truly random data form today's computing
machines (usually PC's). Some ad-hoc methods exist 
\cite{Davis, rfc1750, urandom}
to extract a handful of random bits per second, but the only 
way to get a lot of random bits of guaranteed quality in a short time is
to add a non-deterministic random bit generator to the computer, 
thus effectively realizing the probabilistic Turing machine.
\\

 Problem, however, is to construct a good enough generator, whose output can be
for all practical purposes considered as being truly unpredictable and random.
\\

We describe, in this article, a theory and one practical realization
of a reliable and fast 
random bit generator which should not be expensive to produce.

\section{The method}
\
\label{sect:method}

Since it is obvious that only physical reality may provide
true randomness, the
non-deterministic RBG should be 
based on a repeated measurements of identical, independent physics 
processes and associated method for
extraction of random bits from these measurements.
An RBG should be characterized by the following:

\begin{enumerate}
\item its output can not be predicted regardless of the amount of 
knowledge about it
(just as it would be impossible to predict outcome of flipping a 
fair coin by knowing exactly how it looks);
\item two identical generators may not be synchronized to produce 
the same sequence of bits.
\end{enumerate}
 
 The second requirement could well be regarded as the definition of 
a non-deterministic generator. This requirement may be reformulated so that 
a non-deterministic generator may not accept an initial state. This is
opposite to pseudo-random generators which must accept 
an initial state.
\\

Since we do not expect that measuring of independent physics processes,
that is making independent identical experiments whose outcome is random,
may form any pseudo-random pattern we are left with only two 
problems associated with non-deterministic random bit generators.
These are:
\begin{enumerate}
\item {
       statistical bias, defined as:
       \begin{eqnarray}
        b = p(1) - 0.5
       \label{eq:bias}
       \end{eqnarray}
       where $p(1)$ stands for probability of ones;
      }
\item {
       correlations among bits, among which the most important {\em serial correlation coefficient} is defined as \cite{Knuth}:
       \begin{eqnarray}
        a & = & \frac{n \sum_{i=0}^{n-1}b_i b_{(i+1)\, {\rm mod}\, n} - \left(\sum_{i=0}^{n-1}b_i\right)^2}{ n \left(\sum_{i=0}^{n-1}b_i^2\right) - \left(\sum_{i=0}^{n-1}b_i\right)^2}
       \label{eq:auto}
       \end{eqnarray} 
       where $b_1, i=0 \ldots n-1$ denotes $i$-th bit in the sequence.
      }
\end{enumerate}
Both are measures of imperfections that are inevitable in practical
realizations of generators.
Correlations appear
if the measured physics processes (experiments) are not completely 
statistically independent of each other, 
whereas statistical bias is mainly associated with imperfections in the
measuring equipment (eg. electronics).
Serial correlation coefficient is a measure of the extent 
to which a bit (in a sequence) depends on a previous bit (\cite{Knuth}),
and takes on a value between -1 and 1. 
We suppose that correlations between
bits further apart (corresponding to experiments further apart in time)
are smaller if not negligible.
For truly random sequences of bits, of course, 
both statistical bias and correlations 
tend to zero, when the length of the sequence goes to infinity.
\\

The method presented here solves both problems. 
It consists of counting events which are result of 
measurement of some random physical process. 
For example events may be a radioactive decay or taping of rain drops 
on a tin roof.
These events appear at random but
measured for a long time they have some mean period, $\mu$. The
state of the counter is examined at regular time intervals
with the period $T \gg \mu$. If the counter state is found to be  odd
then the output from the generator is "1", otherwise it is "0".
Unlike some other methods which require exponential distribution 
of time intervals between events \cite{vincent70}
or sampling of an analog white noise source \cite{bucci01},
we will show that this method, in the limit of slow sampling 
($T / \mu \rightarrow \infty$) leads to vanishing bias and correlations 
regardless of the distribution of the random process being measured.
Therefore it may used 
to extract random bits from a large variety of processes.

\section{Practical realization of a random bit generator}
\
\label{sect:realization}

Useful non-deterministic random bit generator built in
hardware should satisfy several requirements:

\begin{enumerate}
\item Bit sampling method must not rely strongly on any property of the
      measured physical process other than its randomness, at least
      in some easily achievable limit;

\item Generator should withstand reasonable tolerance in components 
      and operating conditions (eg. supply voltage, temperature, EM noise)
      without the need for (re)calibration or compensation;

\item Possible malfunctions during the lifetime of the generator
      should be foreseen and checked for at each generator restart
      (for example like recommended in \cite{FIPS140-1}) or even continuously;

\item Sequences of bits produced by the generator should pass,
      with a high probability, any known statistical randomness test.
\end{enumerate}

In the generator described here physical processes 
in Zener diode serve as a source 
of randomness. The noise voltage at terminals of the diode is "measured" 
by a special electronic circuit at regular time intervals.
Each measurement results in a random bit.

\begin{figure}[b]
\centerline{\includegraphics[width=85 mm,angle=0]{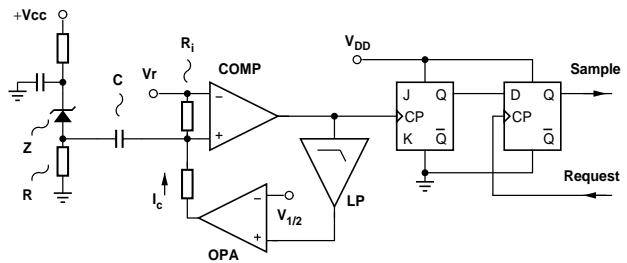}}
\caption{Schematic diagram of the non-deterministic random bit generator}
\label{Fig1}
\end{figure}

The circuit presented in Fig. \ref{Fig1} follows the method 
of operation described in the previous section. 
It consists of the following five blocks:

\begin{enumerate}
\item
 A source of electric noise;
\item
 a DC decoupling capacitor $C$;
\item
 a circuit for digitizing the noise voltage 
   consisting of a comparator controlled with a rough 
   automatic zero-bias correction circuitry;
\item
 a counting circuit (JK flip-flop);
\item
 a sampling circuit which delivers a bit upon an external request.
\end{enumerate}

It is a well known fact that a Zener diode operating 
in a reverse polarity and the current strength near the knee, can serve as 
a noise generator.
For example, a 6.8 Volt commercial Zener diode can produce a noise 
fluctuation of an amplitude of 30 to 50 mV (peak to peak) with a mean 
frequency of zero crossings of the order of 10 MHz \cite{Somlo75}.
\\

It is also well known that a Zener diode with a knee voltage of less 
than 6.2 V operates mainly in the Quantum Mechanical (tunneling) regime, 
while the diodes with the knee above that operate mainly in the micro 
plasma regime \cite{Somlo75}. Both regimes have ideal properties of 
unpredictability needed for a truly random noise voltage source.
\\

The best temperature stability of the noise amplitude is also obtained 
in the Zener diodes with the knee voltage of approx. 6.2 V, 
because in this case the two regimes with opposite temperature coefficients 
are in equilibrium \cite{Somlo75}. The same condition is also optimal 
from the point of the long term stability.
\\

The generator functions in the following way.
Noisy voltage from the Zener diode is AC decoupled from the digitizing stage 
which consists of a comparator COMP controlled with a rough automatic 
zero-bias feedback control. The role of the comparator
is to convert tiny analog noise to the digital signal suitable
for further processing.
\\

The capacitor $C$ in the series with the output resistance $R$ of the noise
source and the effective input resistance $R_i$ of the comparator
forms a major contribution to an unwanted "memory". The timely persistence
$\tau_e$ of this memory is equal to:
\begin{eqnarray}
 \tau_e = C (R + R_i).
\label{eq:taue}
\end{eqnarray}
Voltage amplitudes of any two noise
variations which happen within the period $\tau_e$ will
be mutually correlated because
of the electric charge in the capacitor $C$
which has no time to discharge through
the resistance in the system. 
If physical events are not completely independent of each other there
will be another persistence $\tau_p$ giving rise to 
the total memory:
\begin{eqnarray}
 \tau = \tau_e + \tau_p
\label{eq:tau}
\end{eqnarray}
Luckily, this
"memory" effect dies off exponentially with the 
time distance between the two variations,
thus one can conclude that any two variations that are distant enough
in time may  be considered as statistically independent. 
Whenever this applies, our method is valid, as will be explained later.
Nevertheless, the "memory" of the circuit limits frequency 
bandwidth of the noise and sets an absolute upper 
limit to the bit extraction rate from the generator, 
which limit is independent of the latter bit extraction method.
\\

\begin{figure}[b]
\centerline{\includegraphics[width=85 mm,angle=0]{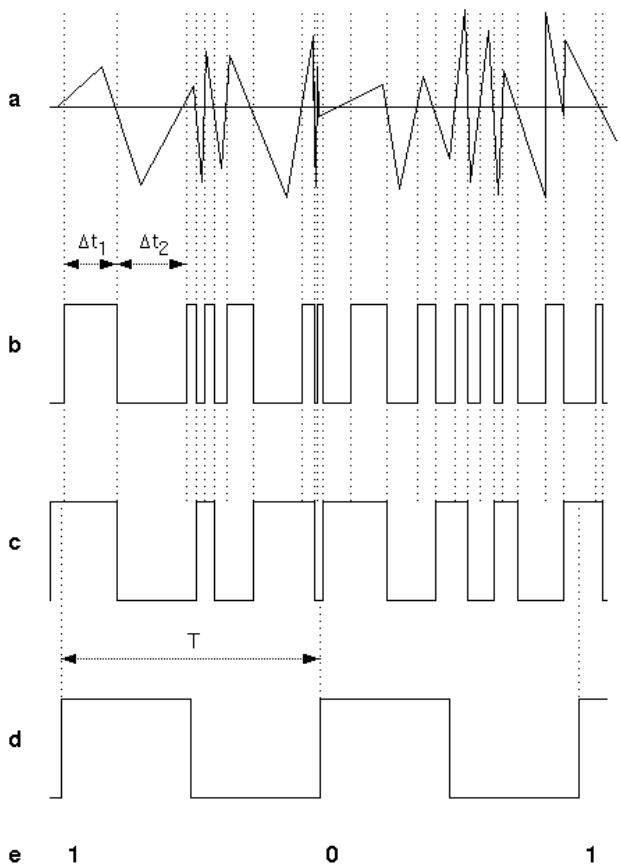}}
\caption{Signals at different points of the hardware generator: a) analog diode noise, b) output of the comparator COMP, c) output Q of the JK flip-flop, d) final output of the generator (D flip-flop).}
\label{Fig2}
\end{figure}

The negative input of the comparator COMP is connected to a suitable
DC reference voltage $V_r$. 
 Between the two inputs of the comparator (positive and negative) a small 
DC "offset" voltage can be induced by virtue of the resistor $R_i$ and 
the control current $I_c$ that flows through it.
As the result, the positive input of the comparator "sees" the sum of the
offset voltage and the noise voltage. Whenever the sum exceeds the 
reference voltage $V_r$, the output of the comparator goes into the high 
logical state "1", whereas when the sum goes below the $V_r$, 
the output goes into the low logical state "0". 
This is illustrated in the Fig. \ref{Fig2}
\\

By setting the control current $I_c$ a little lower or higher,
the output of the comparator COMP spends respectively a little more or less 
time in the logical state "1".
At certain strength of the control current $I_c$, 
one can achieve that the comparator
COMP spends approximately equal amount of time in either 
of the logical states, 
the point at which the comparator has the highest efficiency of 
detecting voltage variations.
In other words, the duty cycle of the comparator COMP would then on average 
be close to 0.5. 
This is ensured by setting $V_{1/2} = \frac{1}{2} V_{DD}$.
If, for any reason, the duty cycle ever gets changed, the feedback 
network closed by comparator COMP, operational amplifier OPA and 
low-pass filter LP (Fig. \ref{Fig1}) 
will respond by changing the control current $I_c$ so
that the duty cycle of 0.5 will be restored.
This behavior effectively solves the requirement 2.
Already here, at the output of the comparator COMP, there is an 
approximately equal chance that the state is at logical "0" or 
at logical "1", i.e. the bias of the sampled output would be
 close to zero.
However, due to technical reasons it is quite difficult to keep the bias 
below 1/1000 for long term and even this would be not possible
without some fine-tuning.
Tuning the bias consumes a lot of time (due to its statistical
nature) and would therefore be unfavorable for a mass production 
of the generator.
\\

 Present method for extracting random bits 
eliminates the need to tune the bias to zero value, 
and actually allows to achieve bias as low as desired, 
without the need to do any modifications to the circuit. 
\\

Event counting is the crucial point of the method.
Namely, the JK-type flip-flop (Fig. \ref{Fig1}) 
performs a continued counting of the events,
but keeps track only whether the count is even or odd.
The result of the counting appears at the output Q of the JK flip-flop 
and represents a new random bit sequence with highly suppressed bias and 
serial correlation.
To understand how this works it is important to keep in mind that the 
output Q of the JK flip-flop shall be sampled periodically in time.
This is a condition for good operation of this generator. 
The sampling is done by the D-type flip-flop.
\\

Without the loss of generality, let us suppose that at the time zero 
($t = 0$) the Zener diode Z exhibits a voltage breakdown. 
(Voltage breakdown is only one of the processes that may cause fluctuations of the voltage across the diode that are manifested as noise. But to simplify the language we will refer to "voltage breakdown" as a synonym for a sudden positive jump of the voltage across the Zener diode.) 
This causes the output of the comparator COMP to exhibit a positive going transition (and later a negative one). 
Every such event triggers the JK flip-flop to reverse 
its state at the output Q. 
This is illustrated in the Fig. \ref{Fig2}.b and \ref{Fig2}.c. 

Let us suppose that the voltage breakdowns happen at times 
$t_0$ = 0, $t_1$, $t_2$, $t_3$, $t_4$, etc.
Tiny intervals of time between neighboring voltage breakdowns 
$\Delta t_k = t_k - t_{k-1}$ where $k$ = 1, 2, 3...
are distributed according to some statistical distribution. 
The mean value $\mu$  of duration of this tiny intervals is defined as: 
\begin{eqnarray}
  \mu = \lim_{N \rightarrow \infty} \frac{1}{N} \sum_{k=1}^{N} t_k
\label{eq:inter}
\end{eqnarray}

For example, in the pure Quantum Mechanical regime of the Zener diode, 
intervals between neighboring voltage breakdowns $\Delta t_k$ follow the 
Exponential distribution. 
Mixing of the Quantum Mechanical model with the plasma noise, 
noise sources other than a Zener diode or other models including effective 
memory and/or filtering of the noise signal prior to digitization, 
may lead to more centered distributions such as Poisson or even Uniform-like 
distribution. 
The power of this method lies in the fact that the knowledge of the actual 
distribution of the time intervals between neighboring voltage breakdowns 
is irrelevant. The only important thing is that the shape of the distribution
stays stable over a period of time substantially 
larger than the sampling period $T$.
\\

The method presented in this work consists in sampling 
(i.e. reading out) of the 
output Q of the JK flip-flop periodically, at times 
$T$, $2 T$, $3 T \ldots$
and generally at times 
$i \times T$ with $i = 1, 2, 3 \ldots$ .
The sampling is made by a D-type flip-flop (Fig. \ref{Fig1}). Namely, when a positive-going edge of the sampling signal (that is transition from "0" to "1") appears at the "Request" input, state that is present at the input D gets sampled. Sampled value is memorized (frozen) and displayed at the output Q of the D-type flip-flop, and stays there until a new positive-going edge appears. According to the definition of the D-type flip-flop, the sampling is done almost instantaneously so its output stays frozen for almost the whole sampling period $T$ which makes possible for other devices (such as a computer) to read it safely. 
\\

Let us now suppose that the sampling period $T$ is chosen large enough so 
that many voltage breakdowns occur during each sampling period. 
In another words, we suppose: 

\begin{eqnarray}
T  \gg  \mu.
\label{eq:tggmu}
\end{eqnarray}
 
This situation is illustrated in the Fig. \ref{Fig2} 
where several voltage breakdowns in the noise signal (FIG.2.a) 
happen during any sampling period of length $T$ (FIG 2.d).

Furthermore, let us denote that the JK flip-flop changes its state at the 
output Q: 
$N_0$ times from $t = 0$ till $t = T$,
$N_1$ times from $t = T$ till $t = 2 T$, 
and generally $N_i$ times from $t = i T$, till $t = (i + 1) T$, 
where $i = 1, 2, 3, \ldots$ .
\\

Now we can conclude that every sampling period (of length $T$)
approximately equals the sum 
of many tiny intervals $\Delta t_k$:

\begin{eqnarray}
 T & \approx & \sum_{k=1}^{N_0} \Delta t_k \nonumber \\
 T & \approx & \sum_{k=N_0+1}^{N_0+N_1} \Delta t_k \nonumber \\
 {\rm etc.} & & 
\label{eq:sums}
\end{eqnarray}

The above equalities hold only approximately because the first and the last of the tiny intervals in a given sum may be only partially happening during the respective sampling period $T$. 
In such a case, a part of the first tiny interval may have actually happened during the previous period, while a part of the last tiny interval may have actually happened during the next period (sum). However, approximation can be made arbitrarily good by setting the large enough sampling period $T$.
\\

The Central limit theorem of the Statistics states that the sum of a large number (say $N$) of independent random variables (for example $\Delta t_k$), which follow some (any) distribution,  approaches Normal distribution in the limit of large $N$. 
This means that the statistical variable $x$ defined as:
\begin{eqnarray}
   x = \sum_{k=1}^{N} \Delta t_k,
\label{eq:xsum}
\end{eqnarray}

follows approximately the Normal distribution for large $N$. 
However, in our case, defined by equations (\ref{eq:sums}), the "variable" $x$ 
is fixed (being just $T$), and the relevant statistical random 
variable becomes the number of summands, $N$ defined as:

\begin{eqnarray}
   \sum_{k=1}^{N} \Delta t_k \leq
   x < \sum_{k=1}^{N+1} \Delta t_k.
\label{eq:Ndef}
\end{eqnarray}
  
In the special case mentioned above, when the tiny 
intervals $\Delta t_k$ are distributed according to the 
Exponential distribution, random variable $N$ follows, 
by definition, the Poisson distribution:

\begin{eqnarray}
 P(N) = \frac{(T/\mu)^N}{N!} exp(-T/\mu).
\label{eq:poisson}
\end{eqnarray}

Quite generally, regardless of the distribution function of tiny intervals 
$\Delta t_k$, it can be shown \cite{Stip} that the integer random variable $N$ is distributed according to the Binomial distribution $B(N; 2T/\mu, p)$, where $p$ quickly approaches 0.5 as $T / \mu$ tends to infinity. 
Given the general expression for the Binomial distribution:

\begin{eqnarray}
 B(r; n,p) = \left( \begin{tabular}{c} $n$ \\ $r$ \end{tabular} \right) p^r (1-p)^{n-r} .
\label{eq:binom}
\end{eqnarray}

distribution of the variable $N$ can be written as:

\begin{eqnarray}
B(N;2T/\mu,p) = \left( \begin{tabular}{c} $2T/\mu$ \\ $N$ \end{tabular} \right) p^N (1-p)^{2T/\mu-N}. 
\label{eq:binom2}
\end{eqnarray}

This is not in a contradiction with the previous result because the 
Poisson distribution (\ref{eq:poisson}) 
becomes equal to the Binomial distribution
(\ref{eq:binom2})
in the limit $T / \mu \rightarrow \infty$.
\\

Even without a rigorous proof it is easy to understand why $N$ follows symmetric Binomial distribution (\ref{eq:binom2}) in the limit of slow sampling or, 
equivalently, large $N$. 
Following eq. (\ref{eq:Ndef}), for large $N$ we can write:
\begin{eqnarray}
   x \approx \sum_{k=1}^{N} \Delta t_k.
\label{eq:xapprox}
\end{eqnarray}
Approximating the definition (\ref{eq:inter}) with:
\begin{eqnarray}
 \mu \approx \frac{1}{N} \sum_{k=1}^{N} \Delta t_k
\label{eq:mean}
\end{eqnarray}
to eliminate the sum in (\ref{eq:xapprox}) we may conclude:
\begin{eqnarray}
 N \approx \frac{1}{\mu} x .
\label{eq:explanation}
\end{eqnarray}
Since $x$ follows Normal p.d.f. so should $N$. 
But since $N$ is an integer number it actually follows
a discrete version of the Normal p.d.f., i.e. 
symmetric Binomial distribution with $p$=0.5,
{\em Q.E.D}.
This is a very important result because it means that (\ref{eq:binom2}) 
holds regardless of the distribution of tiny intervals $\Delta t_k$.
This solves the requirement 1.
\\

It can be shown \cite{Stip} that as a special consequence of the equation 
(\ref{eq:binom2}) 
the probability that the random variable $N$  is even becomes equal to the probability that it is odd, in the limit $T / \mu \rightarrow \infty$.
But, when the JK flip-flop changes its state an even number of times its output Q returns to the initial state (say "0") whereas when it changes an odd number of times it returns the complementary state (say "1"). 
Thus, in the limit of slow sampling, 
the probability of producing "0" would be equal to the probability of producing "1" at the output Sample (Fig. \ref{Fig1}). This fact can be expressed like this:

\begin{eqnarray}
 T / \mu \rightarrow \infty  \Rightarrow b \rightarrow 0.
\label{eq:biastozero}
\end{eqnarray}

Thus it has been shown that the generator and the corresponding method of 
sampling random phenomena solve the first problem stated in 
the section \ref{sect:method}, 
i.e. that the bias should vanish in some limit.
\\

The second problem, namely the
statistical independence of the logical states (bits) at the output of the generator, is solved as follows. First we realize that the subsequent voltage breakdowns in the Zener diode are statistically independent because the breakdown itself is an unpredictable physical phenomenon. Secondly, the internal capacitive memory that appears in the circuit (which for example may be introduced by imperfections in electronics design, or be a part of physical process governing the alternate current noise source) "dies off" exponentially with the half-life $\tau$ (Eq. (\ref{eq:tau})).
Effect of the memory on correlations can be made as small as desired in the limit of slow sampling, that is by making T/$\tau$ large enough. 
Namely, in the limit 
 $T / \tau \rightarrow \infty$
sums of the type (\ref{eq:sums}) become mutually statistically independent. The statistical independence of two subsequent sums, in that limit, takes place because the two sums contain large portion of summands that are "out of the range" of the memory effect of each other. If we consider two sums that are not subsequent, then the memory effects are even smaller. This means that a bit corresponding to a sum becomes statistically independent of neighboring and even more all other bits, in the limit of slow sampling. 
It can be concluded that the 
generator and the corresponding method of sampling random phenomena also solve the second problem stated in 
the section \ref{sect:method}.
\\
 
Functioning of the whole circuit in Fig. \ref{Fig1} can be best understood by looking at signals shown in the Fig. \ref{Fig2}a-e.
One can see that with each positive-going edge at the output of the comparator COMP (Fig. \ref{Fig2}.b), the JK flip-flop changes state at its output Q (Fig. \ref{Fig2}.c). The sampling signal (Fig. \ref{Fig2}.d) isn't, of course, in any synchronization with these signals. 
Whenever it exhibits a positive-going transition, the output Q of the JK flip-flop gets sampled and the state thus obtained is the output bit from the generator (Fig. \ref{Fig2}.e).
\\

It is important to understand that the output Q of the JK flip-flop always produces "1" after "0" and "0" after "1", therefore its output is not random at all. 
However a duration of zeros and ones varies randomly (Fig. \ref{Fig2}c) 
and periodic sampling of the said output by the D-type flip-flop provides a
good quality random bits, in the limit of slow sampling 
($T \gg \mu$ and $T \gg \tau$). 

\section{Simulations}
\
\label{sect:simulation}
 
In order to check the theory of operation described above,
a series of simulations of the circuit in Fig. \ref{Fig1} were made.
Simulations should provide an insight on how 
the type of distribution of tiny intervals between
subsequent physical events and ratios $T/\tau$ and $T/\mu$ 
influence the quality 
of generated bits.
The circuit was simulated by a specifically designed computer program
which simulates a random digital signal 
obeying a specified distribution which would 
be present at the 
output of the comparator COMP, and assuming that this signal
gets processed by ideal flip-flops. The approximation of considering 
the two flip-flops to be ideal is very good because their 
fall and rise times (of approximately 7 nanoseconds) are
much shorter than the changes in the noise being processed.
\\

Two distributions of tiny intervals were simulated: Exponential and Uniform.
Exponential distribution is natural distribution for many processes. 
Uniform distribution assumes that the intervals between physical
events may be between zero and 1 
(in arbitrary time units, for example microseconds) 
and evenly distributed between the two limits.  
The two distributions are chosen because they are very different: for example
Exponential d. is completely non-symmetric and peaked at one value (zero) 
whereas Uniform d. is symmetric and has no peak, etc.
Most processes should obey distributions that are somewhere between 
those two extremes.
\\

The effect of memory was simulated as a pile up: 
any two breakdowns closer (in time) than
$\tau$ were considered as being just one breakdown happening at the 
end of the last breakdown. 
This in effect may chain up more than two events to be seen as one.
This is a very faithful model for a noise source which we use, and
a reasonable approximation for any other noise source because memory
(by definition) always piles up events in a way that events which
happen to close to one another get piled up and be
detected as just one event.  
\\

The method for extracting bits does not prefer zeros nor ones,
therefore we do not expect the bias to be significant even in the case 
when both ratios
$T/\tau$ and $T/\mu$ are small. What  happens however is that when 
sampling is being made too fast there is an excess of longer patterns of
ones or zeros. 
This in turn means that the probability of single ones (pattern "010")
a well as that of single zeros (pattern "101") will be smaller than
expected (expected vales are: $p(010) = p(101) = 1/8$).
\\

For each choice of distribution, ratio $T/\tau$ and ratio $\tau/\mu$, 
the program calculates (by simulation) three output values: 
probability of ones $p(1)$, 
probability of single ones $p(010)$ 
and serial correlation coefficient $a$. 
Results of simulations for the two distributions and 
a range of ratios  $T/\tau$ and $\tau/\mu$ are shown in the
tables \ref{Tbl1}-\ref{Tbl6}. Values shown are
bias ($b = p(1)-0.5$), 
second order bias ($b_{010} = p(010) - 0.125$), 
and serial correlation coefficient.
Number of bits generated in each simulation is 6,250,000. According to
that statistics, 1 sigma errors are:
0.00020  for $b$, 
0.00007  for $b_{010}$ and
0.00040  for $a$.
As a source of random numbers the program used semi-hardware 
"{\tt /dev/urandom}" Linux kernel random bit generator.
The result were cross-checked using 
the standard implementation of "secure" software random number 
generator {\em rand()} that is part of the Linux C compiler 
(we used gcc version 2.95).
No discrepancies were found in the results within the error margins.
\\

\begin{table}[h]
\begin{tabular}{l|rrrrrr}
\hline
 & \multicolumn{6}{c}{$\tau / \mu$} \\
$T / \mu$ &  0  &  0.2  &  0.33  &  0.5  &  0.75  &  1   \\
\hline
 2  &  0.00013 &  0.00010 & -0.00007 &  0.00030 & -0.00029 &  0.00004  \\
 3  & -0.00030 &  0.00028 & -0.00010 &  0.00026 & -0.00013 &  0.00041 \\
 4  & -0.00005 &  0.00019 &  0.00022 & -0.00020 & -0.00003 & -0.00030 \\
 6  & -0.00000 &  0.00036 & -0.00015 &  0.00001 &  0.00020 &  0.00013 \\
 9  &  0.00011 & -0.00005 &  0.00012 &  0.00003 & -0.00005 & -0.00029 \\
 12  & -0.00013 &  0.00010 & -0.00013 & -0.00027 &  0.00037 & -0.00013 \\
\hline
\end{tabular}
\caption{Uniform distribution, bias}
\label{Tbl1}
\vspace*{0.2cm}
\begin{tabular}{l|rrrrrr}
\hline
 & \multicolumn{6}{c}{$\tau / \mu$} \\
$T / \mu$ &  0  &  0.2  &  0.33  &  0.5  &  0.75  &  1  \\
\hline
 2  &  0.00955 &  0.00442 &  0.00383 &  0.01479 &  0.06978 &  0.10538 \\
 3  & -0.01351 & -0.01744 & -0.02429 & -0.04041 & -0.05231 & -0.02093 \\
 4  &  0.00701 &  0.01220 &  0.01852 &  0.02598 & -0.00794 & -0.03638 \\
 6  & -0.00100 & -0.00121 & -0.00227 & -0.01113 & -0.01609 &  0.01694 \\
 9  &  0.00028 & -0.00001 &  0.00015 & -0.00224 & -0.00261 &  0.00509 \\
 12  &  0.00006 &  0.00000 &  0.00015 & -0.00038 &  0.00017 & -0.00049 \\
\hline
\end{tabular}
\caption{Uniform distribution, b(010)}
\label{Tbl2}
\vspace*{0.2cm}
\begin{tabular}{l|rrrrrr}
\hline
 & \multicolumn{6}{c}{$\tau / \mu$} \\
$T / \mu$ &  0  &  0.2  &  0.33  &  0.5  &  0.75  &  1  \\
\hline
 2  & -0.05748 & -0.04355 & -0.05233 & -0.11648 & -0.31369 & -0.40496 \\
 3  &  0.06358 &  0.08214 &  0.11607 &  0.20639 &  0.27048 &  0.06288 \\
 4  & -0.03136 & -0.05413 & -0.08151 & -0.11538 &  0.01226 &  0.17930 \\
 6  &  0.00477 &  0.00503 &  0.01013 &  0.05077 &  0.07227 & -0.07653 \\
 9  & -0.00004 &  0.00024 & -0.00057 &  0.00963 &  0.01214 & -0.02286 \\
 12  &  0.00044 &  0.00017 & -0.00077 &  0.00152 & -0.00170 &  0.00254 \\
\hline
\end{tabular}
\caption{Uniform distribution, serial correlation}
\label{Tbl3}
\vspace*{0.2cm}
\end{table}
\begin{table}[h]
\begin{tabular}{l|rrrrrr}
\hline
 & \multicolumn{6}{c}{$\tau / \mu$} \\
$T / \mu$ &  0  &  0.2  &  0.33  &  0.5  &  0.75  &  1  \\
\hline
 2  &  0.00003 &  0.00023 & -0.00025 &  0.00013 &  0.00033 &  0.00005 \\
 3  &  0.00018 & -0.00013 &  0.00003 & -0.00016 &  0.00034 &  0.00006 \\
 4  & -0.00010 & -0.00033 & -0.00013 & -0.00014 & -0.00013 &  0.00005 \\
 6  &  0.00002 & -0.00013 & -0.00013 &  0.00022 & -0.00013 & -0.00016 \\
 9  & -0.00013 &  0.00029 &  0.00014 & -0.00011 & -0.00016 &  0.00018 \\
 12  & -0.00021 & -0.00001 &  0.00002 & -0.00000 &  0.00032 &  0.00024  \\
\hline
\end{tabular}
\caption{Exponential distribution, bias}
\label{Tbl4}
\vspace*{0.2cm}
\begin{tabular}{l|rrrrrr}
\hline
 & \multicolumn{6}{c}{$\tau / \mu$} \\
$T / \mu$ &  0  &  0.2  &  0.33  &  0.5  &  0.75  &  1  \\
\hline
 2  & -0.00398 & -0.00101 &  0.00168 &  0.01385 &  0.04992 &  0.04163 \\
 3  & -0.00080 &  0.00008 & -0.00016 & -0.00299 &  0.00777 &  0.04519 \\
 4  &  0.00016 & -0.00015 &  0.00014 & -0.00013 & -0.00939 &  0.00454 \\
 6  &  0.00008 & -0.00001 &  0.00005 & -0.00011 &  0.00209 & -0.00742 \\
 9  & -0.00017 &  0.00000 & -0.00002 & -0.00007 & -0.00029 &  0.00139 \\
 12  &  0.00002 & -0.00006 & -0.00003 & -0.00008 & -0.00020 & -0.00045 \\
\hline
\end{tabular}
\caption{Exponential distribution, b(010)}
\label{Tbl5}
\vspace*{0.2cm}
\begin{tabular}{l|rrrrrr}
\hline
 & \multicolumn{6}{c}{$\tau / \mu$} \\
$T / \mu$ &  0  &  0.2  &  0.33  &  0.5  &  0.75  &  1  \\
\hline
 2  &  0.01869 &  0.00423 & -0.00729 & -0.06256 & -0.20970 & -0.20169 \\
 3  &  0.00306 &  0.00016 &  0.00042 &  0.01389 & -0.03947 & -0.19117 \\
 4  & -0.00001 &  0.00033 & -0.00016 &  0.00084 &  0.04353 & -0.02615 \\
 6  & -0.00058 & -0.00029 & -0.00023 &  0.00021 & -0.00972 &  0.03569 \\
 9  &  0.00052 & -0.00021 &  0.00020 &  0.00022 &  0.00095 & -0.00623 \\
 12  &  0.00033 &  0.00032 &  0.00021 &  0.00006 &  0.00015 &  0.00125 \\
\hline
\end{tabular}
\caption{Exponential distribution, serial correlation}
\label{Tbl6}
\vspace*{0.2cm}
\end{table}

As expected, bias is always consistent with zero.
Absolute values of the second order bias and 
the serial correlation coefficient quickly drop to zero
(within the error margins)
when the sampling period becomes large enough and memory effects small enough.
The Exponential distribution seems slightly more favorable in this respect 
than the Uniform.
From these results it can be concluded that in the worst case
good bits are achieved
by generators with 
the memory effect 
$\tau/\mu < 0.33$
and the ratio $T/\mu > 9$.
This means that
using a standard Zener diode with 3 MHz mean frequency of voltage turnovers
it should be possible to build a generator that is capable of
generating up to several hundred kilobits of good-quality 
random bits per second.

\section{Prototype testing}
\
\label{sect:testing}

We have constructed a prototype based on the block-diagram 
in Fig. \ref{Fig1} using a Zener diode, standard analog chips and CMOS logic.
Bits were produced at 300 kbit/sec. We built four identical circuits
and fed their outputs to a PC computer, thus obtaining
a total of 1.2 Mbit/sec. Thanks to such a high speed we were able
to easily produce long sequences of bits for the subsequent testing.
The sequences of bits produced by the physical generator were tested by three
batteries of statistical tests: Walker's ENT \cite{ENT},
Marsaglia's DIEHARD \cite{DIEHARD} and 
NIST's Statistical Test Suite \cite{sts-1.50}.
\\

The ENT consists of battery of standard tests: entropy, Chi-square,
mean value, Monte Carlo pi value test and the serial correlation test.
These tests may be evaluated for short sequences thus making possible
a fast check of the prototype in the development stage.
The most important of them are the mean value 
and the serial correlation test because
they best grasp problems in hardware.
The mean value addresses the inequality
between number of ones and zeros which is very hard to keep close to zero
with imperfect hardware. 
Non zero serial correlation reflects existence of a short-term memory 
in the system, as
explained in the section \ref{sect:realization}.
Indeed, at the development stage of the prototype  a slight consistent
bias (positive, of the order of $10^{-4}$) has been noticed 
on all four gadgets 
which was attributed to the fact that 
comparator had quite a different rise time from the fall time at its
output. This was corrected by taking another comparator after which the
bias became undetectable.
\\

Once the generator was debugged and ENT tests passed well,
we proceeded by generating long sequences (approx. 10$^8$ bits) 
needed by
the DIEHARD, probably the stringiest 
battery of tests known today. 
Table \ref{tb:diehard} shows a typical 
result of testing of $8\times10^7$ bits (10 megabyte) of random data 
obtained from the generator.
\\

\begin{table}[h]
\vspace*{0.3cm}
\begin{tabular}{l|c}
Test  name &p-value \\
\hline
 BIRTHDAY SPACINGS         & \hspace*{2mm}0.6562 \\
 OVERLAPPING 5-PERMUTATION & 0.4666 \\
 OVERLAPPING 5-PERMUTATION & 0.5396 \\
 BINARY RANK 31x31         & 0.4256 \\
 BINARY RANK 32x32         & 0.0246 \\
 BINARY RANK 6x8           & 0.1513 \\
 BITSTREAM TEST            & 0.5340 \\
 OPSO                      & 0.2728 \\
 OQSO                      & 0.6959 \\
 DNA                       & 0.8619 \\
 COUNT THE 1's             & 0.6046 \\
 COUNT THE 1's             & 0.3398 \\
 PARKING LOT               & 0.7513 \\
 MINIMUM DISTANCE          & 0.2088 \\
 3D SPHERES                & 0.2800 \\
 SQUEEZE                   & 0.9366 \\
 OVERLAPPING SUMS          & 0.9472 \\
 RUNS UP                   & 0.3664 \\
 RUNS DOWN                 & 0.8111 \\
 RUNS UP                   & 0.4229 \\
 RUNS DOWN                 & 0.5369 \\
 CRAPS (wins)              & 0.2463 \\
 CRAPS (throws)            & 0.1541 \\
\hline
\end{tabular}
\caption{Typical result of the DIEHARD tests for the hardware non-deterministic generator}
\label{tb:diehard}
\end{table}

According to Marsaglia \cite{DIEHARD} sequence has failed a test if 
the p-value of that test is  
very close to 1 ($p > 0.999999$), otherwise the sequence has passed the test. 
There are 15 different tests in the DIEHARD, but some are run more
than once with different parameters (eg. binary rank test) or
on independent parts of the sequence under test (eg. count the 1's).
Results in the table \ref{tb:diehard}
indicate a random sequence that has passed all the tests.
\\

It is interesting to note that efforts to standardize requirements
for random bit generators for use with cryptographic software 
have been taken, notably the recommendations of NIST \cite{FIPS140-1}.
A sequence of 10$^9$ bits divided in 100 equally long pieces (files) 
has been subjected to the NIST's
battery of randomness tests, version sts-1.50 \cite{sts-1.50}.
We used the recommended test parameters.
Results are summarized in the 
Table \ref{tb:sts150}.
\\

\begin{table}[h]
\vspace*{0.3cm}
\begin{tabular}{l|c|c|l}
Test\# &p-value &Pass rate &Statistical test	\\
\hline
1.     &\hspace*{2mm}0.9240\hspace*{2mm} &\hspace*{2mm}0.9900\hspace*{2mm} &Frequency		\\
2.     &0.8165  &0.9900   &Block-Frequency	\\
3.     &0.8343  &0.9900   &Cusum		\\
4.     &0.0456  &1.0000   &Runs		\\
5.     &0.3041  &0.9900   &Long-Run		\\
6.     &0.7197  &0.9900   &Rank		\\
7.     &0.3504  &0.9900   &FFT		\\
8.     &0.0909  &1.0000   &Aperiodic-Template	\\
9.     &0.7791  &0.9900   &Periodic-Template	\\
10.    &0.2022  &0.9900   &Universal		\\
11.    &0.9240  &0.9800   &Apen		\\
12.    &0.8623  &1.0000   &Random-Excursion	\\
13.    &0.7727  &0.9844   &Random-Excursion-V	\\
14.    &0.8831  &0.9900   &Serial		\\
15.    &0.2022  &0.9900   &Lempel-Ziv		\\
16.    &0.8165  &0.9900   &Linear-Complexity  \\
\hline
\end{tabular}
\caption{Test of randomness extracted from heartbeats}
\label{tb:sts150}
\end{table}

According to the NIST analysis program,
the minimum pass rate for each statistical test with the 
exception of the random
excursion (variant) test is approximately 0.9602 whereas  
the minimum pass rate for the random excursion (variant) 
test is approximately 0.9527.
Taking this into account 
we conclude that all 16 tests were passed with excellent marks.
Passing all three batteries of tests shows that this generator
conforms also to the general requirement 4 stated in the section
\ref{sect:realization}.
\\

As one more check of the sampling method itself bits were generated
from a cardiogram
data of a healthy patient. A record of 1190 consequent 
heartbeats was made in course of an ergometric stress test.
During the test patient is required to walk on a treadmill
whose speed varies according to a standardized procedure.
Test lasts approximately 12 minutes.
The times at which main heartbeat peaks occur were 
considered as physical "events". 
Taking $T/\mu \approx 5$ we obtained a sequence of 
216 bits using our method. Results of ENT
test shown in the table \ref{tb:srce} indicate excellent randomness.
This demonstrates the power of the method which is able
to extract good random bits form a sequence of heartbeats which are 
nearly periodic but have only a slight random jitter.

\begin{table}[h]
\vspace*{0.3cm}
\begin{tabular}{l|c|c|c}
Bit entropy & Chisquare      & Mean value & Serial correlation \\ 
\hline
  0.9977  &  $0.67 / 50$\%  &  $0.528\pm0.034   $  & $ -0.003\pm0.069  $ 
\end{tabular}
\caption{Test of randomness extracted from heart beats}
\label{tb:srce}
\end{table}

The conclusion is that heartbeat timings may very well be used as a
base for a non-deterministic random generator !

\section{Conclusion}
\ 
\label{sect:conclusion}

A method for extraction of non-deterministic random bits from 
random physics processes (such as nuclei decay or noise voltage variations)
is presented.
The method is characterized by the fact that
it ensures 
a good quality random bits in an easily obtainable limit, that is 
the limit of slow sampling, regardless of the distribution of times
between adjacent processes.
A physical random bit generator
which makes use of 
the method and 
Zener diode noise was built and successfully tested.
The generator is shown to conform to general requirements
for non-deterministic generators stated in section \ref{sect:realization}.
A sample of approximately $10^9$ bits produced 
by the physical generator prototype
is subjected to a series of well-known
statistical tests showing no weaknesses.

\section{Acknowledgments}
\

The heartbeat recording was provided by
courtesy of M. Martinis from Rudjer Bo\v skovi\' c Institute and
the Institute for cardiac deseases, Zagreb.
These data were taken with approval of the patient.
\\

Parts of the method and apparatus for producing random bits are subjected to a patent procedure.


\begin{thebibliography}{100}

\bibitem{QCbook}
M.A. Nielsen, I.L. Chuang, 
Quantum Computation and Quantum Information, 
(Cambridge University Press, Cambridge, 2000) p. 2.

\bibitem{Wikipedia-pTuring}
Wikipedia, the free encyclopedia, Internet URL: http://www.wikipedia.org/wiki/Probabilistic\_Turing\_Machine

\bibitem{CRC}
M. J. Atallah, Algorithms and Theory of Computation Handbook, (CRC Press LLC, Boca Raton, 1998) p. 29.

\bibitem{NRinC} 
       W. H. Press, B. P. Flannery, S. A. Teukolsky,
       W. T.  Vetterling, 
       Numerical Recipes in C: The Art of Scientific Computing,
       (Cambridge  University Press, New  York,  1992)

\bibitem{Knuth}
D. E. Knuth, The art of computer programming, Vol. 2, Third edition, 
(Addison-Wesley, Reading, 1997)

\bibitem{RANLUX}
M. Luescher, Comp. Phys. Comm, {\bf 79}(1994) 100.

\bibitem{RANLUX2}
F. James, Comp. Phys. Comm. {\bf 79}(1994) 111.

\bibitem{Hellekalek}
P. Hellekalek, Mathematics \& Computers in Simulation {\bf 46}(1998) 485.

\bibitem{rreview}
W. E. Sharp, C. Bays, Computers \& Geosciences, {\bf 18}(1992) 79.

\bibitem{Miller}
A.J. Miller, P. Mars, Mathematics \& Computers in Simulation, {\bf 19}(1997) 198.

\bibitem{Marsag90}
G. Marsaglia, Statistics \& Probability Letters, {\bf 9}(1990) 345.

\bibitem{Geant}
S. Agostinelli et. al, Nucl. Instr. and Meth. {\bf A} 506(2003) 250.

\bibitem{primality}
M. O. Rabin, J. Number Th. 12(1980) 128.

\bibitem{BB84}
C. H. Bennet, G. Brassard, 
Quantum cryptography: Public key distribution and coin tossing, 
Proceedings of IEEE International Conference on Computers, 
Systems and Signal Processing, 
(IEEE, New York, 1984) p. 175.

\bibitem{Mau93}
U. Maurer,
IEEE Transactions on Information Theory, {\bf 39} (1993) 733.

\bibitem{rc4}
R. L. Rivest, The RC4 Encryption Algorithm,  RSA Data Security Inc., Mar. 12, 1992.

\bibitem{dobbs96}
I. Goldberg, D. Wagner, Dr. Dobb's, January 1996

\bibitem{Davis}
   D. Davis, R. Ihaka, P. Fenstermacher,
   Cryptographic Randomness from Air Turbulence in Disk
   Drives in: 
   Lecture Notes in Computer Science
   Springer Verlag, Berlin, 1984) p. 839. 

\bibitem{rfc1750}
D. Eastlake, S. Crocker, J. Schiller,
Randomness Recommendations for Security, Internet RFC 1750, December 1994.

\bibitem{urandom}
T. Ts'o, private communication. See also Linux man page {\tt urandom(4)} and 
source code of Linux kernel ver. 2.4.18, file {\tt random.c}


\bibitem{vincent70}
C. H. Vincent, J. Phys. {\bf E} 3(1970) 594.


\bibitem{bucci01}
V. Bagini, M. Bucci, A design of Reliable True Random Number Generator 
for Cryptographic Applications, in: Proceedings of CHES'99 Workshop, 
(Springer Verlag, Berlin ,2000) p. 204.


\bibitem{FIPS140-1}
Security Requirements for Cryptographic Modules, 
Federal Information
Processing Standards Publication 140-1, (FIPS, Gaithersburg, 1994)

\bibitem{Somlo75}
P. I. Somlo, Electronics Letters,  11 (1975) 290.

\bibitem{Stip}
M. Stip\v cevi\' c, 
Some properties of inverted distributions and their application in sampling random phenomena, 
to appear in the Cryptology ePrint Archive, Internet URL: \verb+http://eprint.iacr.org+

\bibitem{ENT}
ENT - A Pseudorandom Number Sequence Test Program, J. Walker, Internet URL: \verb+http://www.fourmilab.ch/random/+

\bibitem{DIEHARD}
DIEHARD battery of stringent randomness tests (various articles and software), G. Marsaglia, Internet URL: \verb+http://stat.fsu.edu/~geo/diehard.html+, also available on CDROM online

\bibitem{sts-1.50}
A. Rukhin et. al., A Statistical Test Suite for Random and Pseudorandom 
Number Generators for Cryptographic Applications, 
NIST Special Publication, (NIST, Gaithersburg, 2001)

\end{thebibliography}
\end{document}